\documentstyle[epsfig,aps]{revtex}
\newcommand{\tvet}{\mbox{\boldmath $t$}}
\newcommand{\Tvet}{\mbox{\boldmath $T$}}
\def\circa#1{\,\raise.3ex\hbox{$#1$\kern-.75em\lower1ex\hbox{$\sim$}}\,}

\newcommand  \f  \varphi
\newcommand \bra {\langle}
\newcommand \ket {\rangle}
\newcommand{\be}{\begin{equation}}
\newcommand{\ee}{\end{equation}}
\newcommand{\ben}{\begin{displaymath}}
\newcommand{\een}{\end{displaymath}}
\newcommand{\ba}{\begin{eqnarray}}
\newcommand{\ea}{\end{eqnarray}}
\newcommand{\ban}{\begin{eqnarray*}}
\newcommand{\ean}{\end{eqnarray*}}
\newcommand{\cro}{\dagger}

\begin{document}
\title{\Large\bf Bloch-Nordsieck violating electroweak corrections\\
to inclusive TeV scale hard processes}
\author{\large Marcello Ciafaloni}
\address{\it Dipartimento di Fisica, Universit\`a di Firenze e
\\ INFN - Sezione di Firenze,
 I-50125 Florence , Italy\\
E-mail: ciafaloni@fi.infn.it}
\author{\large Paolo Ciafaloni}
\address{\it INFN - Sezione di Lecce,
\\Via per Arnesano, I-73100 Lecce, Italy
\\ E-mail: Paolo.Ciafaloni@le.infn.it}
\author{\large Denis Comelli}
\address{\it INFN - Sezione di Ferrara,
\\Via Paradiso 12, I-35131 Ferrara, Italy\\
E-mail: comelli@fe.infn.it}
\maketitle
\vskip0.3cm
\begin{abstract}
We point out that, since the colliders initial states
( $e^+e^-,\;p\, p ,\;p\, \bar{p} $, ... ) carry a definite 
nonabelian flavor, electroweak radiative corrections to inclusive hard cross 
sections at the TeV scale 
are affected by peculiar Bloch-Nordsieck violating double
logs. We recall the setup of soft cancellation theorems, and we
analyze the magnitude of the noncancelling terms in the example of
electron - positron annihilation into hadrons.
\end{abstract}

\vskip1.3cm

Interest in logarithmically enhanced electroweak corrections at NLC
energies has recently arisen \onlinecite{cc1,lecce,kuhn,cc2,fadin},
 after the observation - made by
two of us \cite{cc1} - that double and single \cite{lecce} logarithms of
``soft'' Sudakov type are present and sizeable in fixed angle
fermion-antifermion annihilation processes at the TeV scale.
Such logarithms
 occur because at energies $\sqrt{s}$ much larger than the EW scale
$M_W\approx M_Z\approx M$, 
the latter acts as a cutoff for the collinear and infrared (IR) 
divergences that would   be present in the vanishing
$M$ limit.

The study of Sudakov form factors in the fixed angle, high energy,  regime
where the expansion parameter is $\alpha_W (\log\frac{s}{M^2})^2$
is actually a challenging problem, because one has to investigate the
electroweak theory in a transition region between broken and
unbroken theory, in which two mass parameters, the effective photon
mass and the weak bosons mass $M_W\approx M_Z\approx M$ may both be
important.
Attempts to generalize by various approximation methods the one loop
results to higher orders \onlinecite{kuhn,cc2,fadin} have been made, 
but with somewhat
controversial results\footnote{For instance, Refs. \cite{cc2} and 
\cite{fadin}
agree on the observation that photon and weak boson contributions have
to be considered together to restore the gauge symmetry at very large
energies, but differ in the evaluation of the symmetry breaking
contributions.}. Further study is then needed to fully clarify this
point.

The fact remains, however, that the NLC regime is one in
which the electroweak theory acquires, because of the enlarged phase
space, a full-fledged non abelian structure, therefore
giving rise to novel physical features with respect to the LEP regime.   
In this note we wish to point out the fact that even 
{\it inclusive} TeV
scale cross sections are affected by Sudakov { \it double logs},
due to a lack of cancellation of virtual and real emission electroweak
contributions.

In principle, this lack of cancellation is well known, and is
due to the violation of the Bloch-Nordsieck (BN) theorem \cite{bl} in
nonabelian theories. It was initially pointed out for QCD \cite{dft}, 
where it would imply genuine IR divergences at partonic level, but
was found eventually to have no physical consequences because of the
color averaging of the initial partonic states, forced by the
coupling to colorless hadrons. For instance, non factorized IR
singular terms do occur in Drell-Yan processes
, but at next-to-leading level only, where they are possibly suppressed by 
a Sudakov form factor \cite{mueller}, and eventually turn out to be higher
twist \onlinecite{lrs,col}. 

On the other hand, in the electroweak case the initial state is
{\it fixed} and carries a {\it non abelian
charge}. Therefore, no averaging is possible, so that the  BN
violating terms - though finite because of the symmetry breaking scale $M$-
are unavoidable for any inclusive observable and need
to be carefully computed.

In order to understand this problem, let us start recalling the
structure of soft interactions accompanying a hard SM process, of type
\be\label{eq:1}
\{\alpha^I_1p^I_1,\alpha^I_2p^I_2\}\rightarrow
\{\alpha^F_1p^F_1,\alpha^F_2p^F_2,\dots,\alpha^F_np^F_n\}
\ee
where $\alpha,p$ denote flavor/color and  momentum indices of the initial and
final states, that we collectively denote by $\{\alpha_Ip_I\}$
and $\{\alpha_Fp_F\}$. The S matrix for such a process can be written
as an operator in the soft Hilbert space ${\cal H}_S$, that collects
the states which are almost degenerate with the hard ones, in the
form
\be\label{eq:2}
S={\cal U}^F_{\alpha_F\beta_F}\! (a_s,a^\cro_s)\;\;\;
S^H_{\beta_F\beta_I}\! (p_F,p_I)\;\;\;
{\cal U}^I_{\alpha_I\beta_I}\! (a_s,a^\cro_s)
\ee
where ${\cal U}^F$ and ${\cal U}^I$ are operator functionals of the
soft emission operators $a_s,a^\cro_s$.

Eq. (\ref{eq:2}) is supposed to be of general validity \cite{ciaf},
because it rests
essentially on the separation of long-time interactions (the initial
and final ones described by the ${\cal U}$'s), and the short-time
hard interaction, described by $S^H$. The real problem is to find the
form of the ${\cal U}$'s , which is well known in QED \cite{fk}, has
been widely investigated 
in QCD \cite{cm}, and is under debate in the electroweak case
\onlinecite{cc2,fadin}.
 Their only general property is unitarity in the soft Hilbert
space ${\cal H}_S$, i.e.
\be\label{eq:3}
{\cal U}_{\alpha\beta}{\cal U}^\cro_{\beta\alpha'}=
{\cal U}^\cro_{\alpha\beta}{\cal U}_{\beta\alpha'}=
\delta_{\alpha\alpha'}
\ee

The key cancellation theorem satisfied by eq. (\ref{eq:2}) is due to
Lee, Nauenberg and Kinoshita \cite{kln}, and states that soft
singularities cancel upon summation over initial and final soft states
which are degenerate with the hard ones:
\be
\sum_{i\in \Delta(p_I)}^{f\in \Delta(p_F)}
|\bra f|S|i\ket|^2={\rm Tr}_{{\cal H}_S}
({\cal U}^{I\cro}S^{H\cro}{\cal U}^{F\cro}
{\cal U}^{F} S^H {\cal U}^I)=
{\rm Tr}_{\alpha_I}(S^{H\cro}(p_F,p_I)S^H(p_F,p_I))
\ee
where $\Delta(p_I,p_F)$ denote the sets of such soft states, and we
have used the unitarity property (\ref{eq:3}).

Although general, the KLN theorem is hardly of direct use, because it
involves the sum over the initial degenerate set, which is not
available experimentally. In the QED case, however, there is only an
abelian charge index, so that ${\cal U}_I$ commutes with
$S^{H\cro}S^H$, and cancels out by sum over the final degenerate set only.
This is the BN theorem: observables which are inclusive
over soft final states are infrared safe.

If the theory is non abelian, like QCD or the electroweak one under 
consideration, the BN theorem is generally violated,
because the initial state interaction is not canceled, i.e., by working in
color space, 
\be\label{eq:5}
\sum_{f\in \Delta(p_F)}|\bra f|S|i\ket|^2=
\;_S\bra 0|
{\cal U}^{I\cro}_{\alpha_I\beta'_I}(S^{H\cro}S^H)_{\beta'_I\beta_I}
{\cal U}^I_{\beta_I\alpha_I}|0\ket_S
=(S^{H\cro}S^H)_{\alpha_I\alpha_I}
+\Delta\sigma_{\alpha_I}
\ee 
where the ${\alpha_I}$ indices are not summed over and 
$\Delta\sigma_{\alpha_I}$ is, in general, nonvanishing and IR singular.
Fortunately, in QCD the BN cancellation is essentially recovered
because of two features: (i) the need of initial color averaging,
because hadrons are colorless, and (ii) the commutativity of the
leading order coherent state operators (${\cal U}^l$)
 for any given color indices \cite{cm}:
\be\label{eq:6}
{\cal U}^l={\cal U}^l\! (a_s-a_s^\cro )\quad ,\quad
[{\cal U}^l_{\alpha\beta},{\cal U}^l_{\alpha'\beta'}]=0
\ee
We obtain therefore
\be
\sum_{color}{\cal U}^{l \cro}_{\alpha_I\beta'_I}
(S^{H\cro}S^H)_{\beta'_I\beta_I}
{\cal U}^l_{\beta_I\alpha_I}=
\sum_{color}(S^{H\cro}S^H)_{\beta'_I\beta_I} {\cal U}^l_{\beta_I\alpha_I}
{\cal U}^{l \cro}_{\alpha_I\beta'_I}
=
 Tr_{color}
S^{H\cro}S^H
\ee
thus recovering an infrared safe result (for subleading
features, see Refs. \onlinecite{lrs,col,cm}). 

In the electroweak case, in which
$M$ provides the physical infrared cutoff, there is no way out,
because the initial state is prepared with a fixed non abelian
charge. Therefore eq. (\ref{eq:5}) applies, and double log
corrections $\sim \alpha_W\log^2\frac{s}{M^2}$ 
must affect any
observable associated with a hard process, even  the ones which are
inclusive over final soft bosons. This fact is surprising, because one
would have expected such observables to depend only on energy and on
running couplings, while the double logs represent an
explicit $M$ (infrared cutoff) dependence not yet found before. 

In order to compute the uncanceled double logs, a few preliminary
remarks are in order. First, we assume the underlying process to be
hard, involving a scale much larger than $M$. Therefore the lowest
order soft contributions to $\Delta\sigma$ in eq. (\ref{eq:5})
can be simply described by the
external (initial) line insertions of the eikonal current
\be \label{eq:8}
J_a^\mu=\frac{p_1^\mu}{p_1k}t_1^a+\frac{p_2^\mu}{p_2k}t_2^a
\ee
as depicted in Fig. 1. Secondly, we start from the first non trivial order,
where the effect is present and easily understandable. Then,
the calculation is obviously gauge invariant, because in the hard
(Born) cross section the symmetry is restored, and weak isospin is
conserved. For instance, the weak isospin charge resulting from the
squared insertion current (\ref{eq:8}) in the Feynman gauge and from
Fig. 1 is
\be
(\tvet_1-\tvet_1')\cdot(\tvet_2-\tvet_2')= -(\tvet_1 -
\tvet_1')^2=2 \tvet_1 \cdot \tvet_1'- 
\tvet_1^2-\tvet_1'^{ 2} 
\ee
The last expression, obtained using isospin conservation, is      
identical to the axial gauge result. From 
this form of the charge
it is clear that the $Z_0$ and $\gamma$
contributions cancel out between real and virtual terms, and only the
$W$ contribution remains, which is coupled to left handed fermions only. 
 By adding the obvious eikonal radiation factor, we finally obtain
the following formulas for the corrections to the Born cross sections
$ \sigma_{e^+e^-}$ for the hard process defined by eq. (\ref{eq:1}):
\ba
\Delta \sigma_{e^+e^-}^{RR}&=&0\\
\Delta \sigma_{e^+e^-}^{LL}&=&-\Delta 
\sigma_{e^-\bar{\nu}}^{LL}={\cal A}_W(s)
 \;(\sigma_{e^+\nu}^{LL}-
\sigma^{LL}_{e^+e^-}),\;\;\;\;\;\;\;\;\;{\cal A}_W(s)
=\frac{\alpha_W}{4 \pi}\log^2\frac{s}{M^2}\label{eq:10}
\ea
where $s$ is the c.m. energy squared, $\alpha_W={g^2}/({4\pi})$,
$L,R$ refer to the initial fermions chiralities
and where use has been made of the isospin conservation constraints 
$\sigma^{LL}_{e^+e^-}=\sigma_{\nu\bar{\nu}}^{LL}$,
$\sigma_{e^+\nu}=\sigma_{e^-\bar{\nu}}$. 

Eq.(\ref{eq:10}) provides a rather general result because 
the $W$ coupling is universal and because only the
initial state needs to be specified, so that it applies to various kinds of
hard processes of the type of eq. (\ref{eq:1}). 
It is also clear that the double log cancellation is recovered by summing
over flavors ($\Delta \sigma_{e^+e^-}+\Delta\sigma_{\nu e^+}=0$).
The actual magnitude of the effect is however dependent on the hard
process, because the cross section difference between initial
flavors appears in the r.h.s.  of eq.(\ref{eq:10}).
To be definite, consider the example of $  e^+e^- \to q\bar{q}+X$,
i.e. the total cross section associated to two jets with large
transverse energy\footnote{We are not concerned here with experimental
subtleties, coming from contamination with four jet events coming from
two boson production, or with the competing boson fusion mechanism,
which starts at higher perturbative order.}. The Born amplitude in the
hard symmetric limit is $\sim (g^2 \tvet\cdot\Tvet+
g'^2yY)/s$ where $T$ (and $Y$) denote the quark weak isospin
(hypercharge). By squaring and summing over final flavors, we obtain
the cross section difference ($N_f$ is the number of
families\footnote{The special case of the top quark, requiring a heavy mass
cutoff, was considered in\cite{lecce}, but leads to no important differences
at the double log level we are working.} and $N_c$ is the number of colors):
\be
\sigma_{e^+\nu}^{LL}-\sigma^{LL}_{e^+e^-}
=\frac{N_fN_c}{12 \pi\,s}(\frac{g^4}{8}-\frac{g'^4}{4}Y^2)
\;\;\;\;\;\;\;(Y^2=2\,Y_L^2+Y_R^2+Y_{R '}^2=2\frac{1}{36}+\frac{4}{9}
+\frac{1}{9})
\ee
which occurs in eq.(\ref{eq:10}), and yields

\be
\frac{\Delta \sigma_{e^+e^-}^{LL}}{\sigma^{LL}_{e^+e^-}}\simeq 0.8
\,
{\cal A}_W(s)
\ee

In this case the non canceling terms are positive for initial 
$e^+e^-$ beams
and no particular suppression is noticed.
Given the size of this effect, it is advisable to compute 
higher orders as well.
We think that, due to the inclusive nature of the measurement,               
the QED scale cannot play, in this case, an important role.
It is then tempting to compute higher orders on the  basis of 
the leading form (\ref{eq:6})
of the coherent state operator \cite{cm}
with gauge group $SU(2) \times U(1)$ and only one cutoff, that is  $M$.
With this assumption, a straightforward 
calculation shows that $\gamma$  and $Z$ contributions cancel out 
in the general case as well, and the W contribution exponentiates 
in eq.(\ref{eq:10})
 in the form\cite{new}
\be \label{eq:14}
{\cal A}_W(s) \rightarrow \frac{1}{2}\,(1-e^{-2\,{\cal A}_W(s)})
\ee
This means that, when the energy increases, the noncancellation
in eq. (\ref{eq:14}) becomes maximal and 
the initial state Sudakov effects equalize
eventually the  electron and neutrino beam cross sections.
The universal exponent appearing in eq. (14) is that of the Sudakov 
form factor in the adjoint representation, whose relevance was noticed
for QCD by Mueller \cite{mueller} and by Catani et al \cite{lrs}, and is 
proved for the present case in \cite{new}.

 So far,  we have concentrated on lepton colliders, 
because EW double logs are more directly relevant in such a case.
It is amusing to note that the non canceling
terms affect hadron colliders  as well, because hadrons carry EW charges too.
For instance, at parton level, a formula like (\ref{eq:10})
 holds with an initial quark doublet also, while EW effects
 wash out in the quark-gluon and, of course, in the gluon-gluon cases.
We think therefore that such  EW double logs are to 
be seriously considered for LHC too.

To sum up, TeV scale accelerators open up a regime in which we really
see  non abelian charges at work: even inclusive cross sections have
large electroweak corrections. 
For instance, in our example of $e^+e^-$ into hadrons,
while QCD corrections are O($\frac{\alpha_s}{\pi}$), 
the electroweak ones are  
O($\frac{\alpha_W}{4\pi}\log^2\frac{s}{M^2}$), 
which increases with energy and is
already 7\% at the TeV threshold.

\begin{figure}[htb]\setlength{\unitlength}{1cm}
\begin{picture}(12,7.5)
\put(0.3,1){\epsfig{file=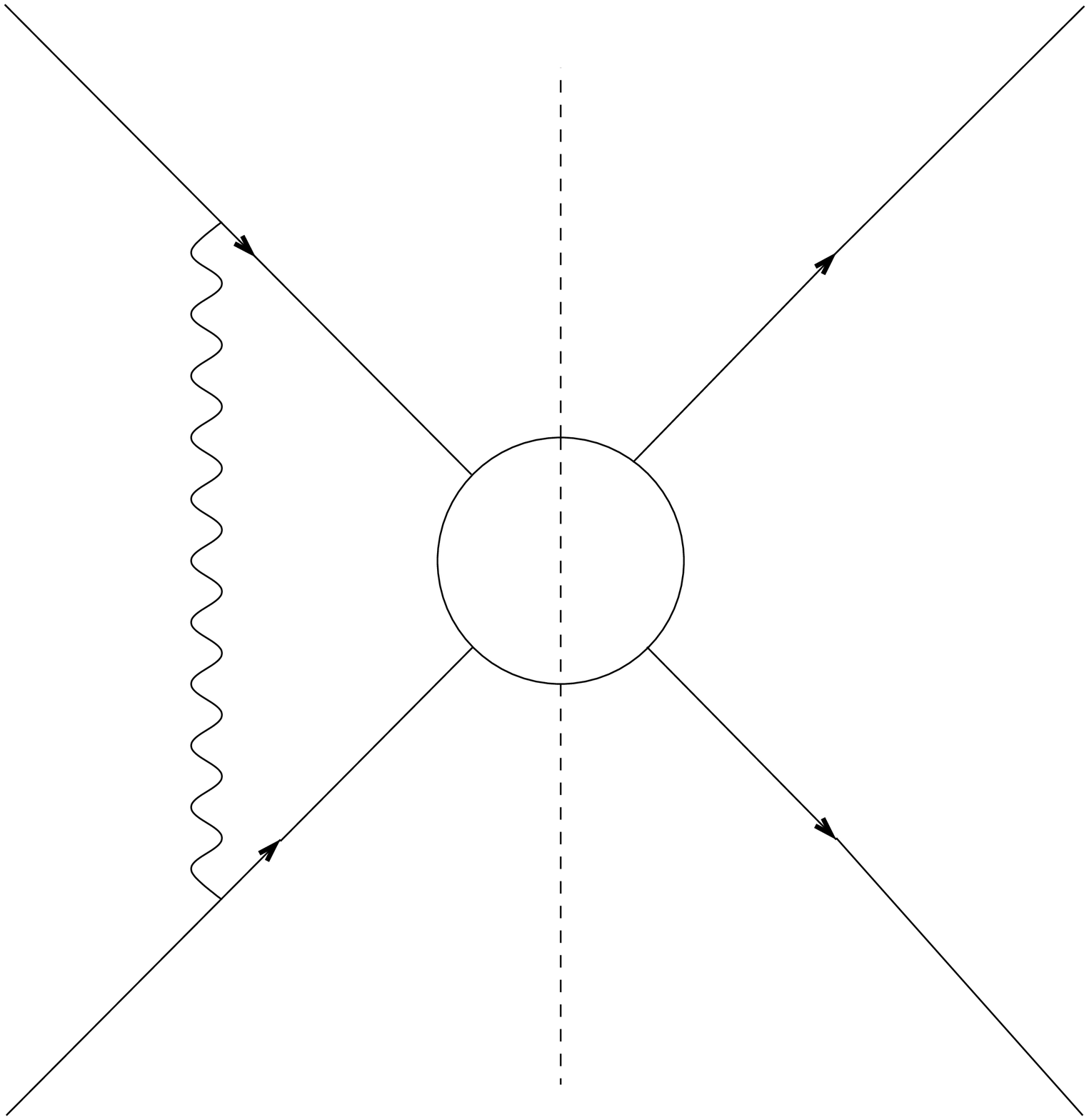,height=6cm}}
\put(.6,4){$k,a$}
\put(.7,6.8){$p_2,\alpha_2$}
\put(.7,0.9){$p_1,\alpha_1$}
\put(3.2,0.52){(a)}\put(12.2,.52){(b)}
\put(10.1,2.2){$t_1^a$}\put(13.7,6.1){$t_2'^a$}
\put(14.3,4){$k,a$}
\put(9,6.8){2}\put(15.3,6.8){2}
\put(9,0.9){1}\put(15.3,0.9){1}
\put(6.4,6.8){2}\put(6.4,0.9){1}
\put(2.8,3.9){$S_H^\cro\,\, S_H$}
\put(9.3,1){\epsfig{file=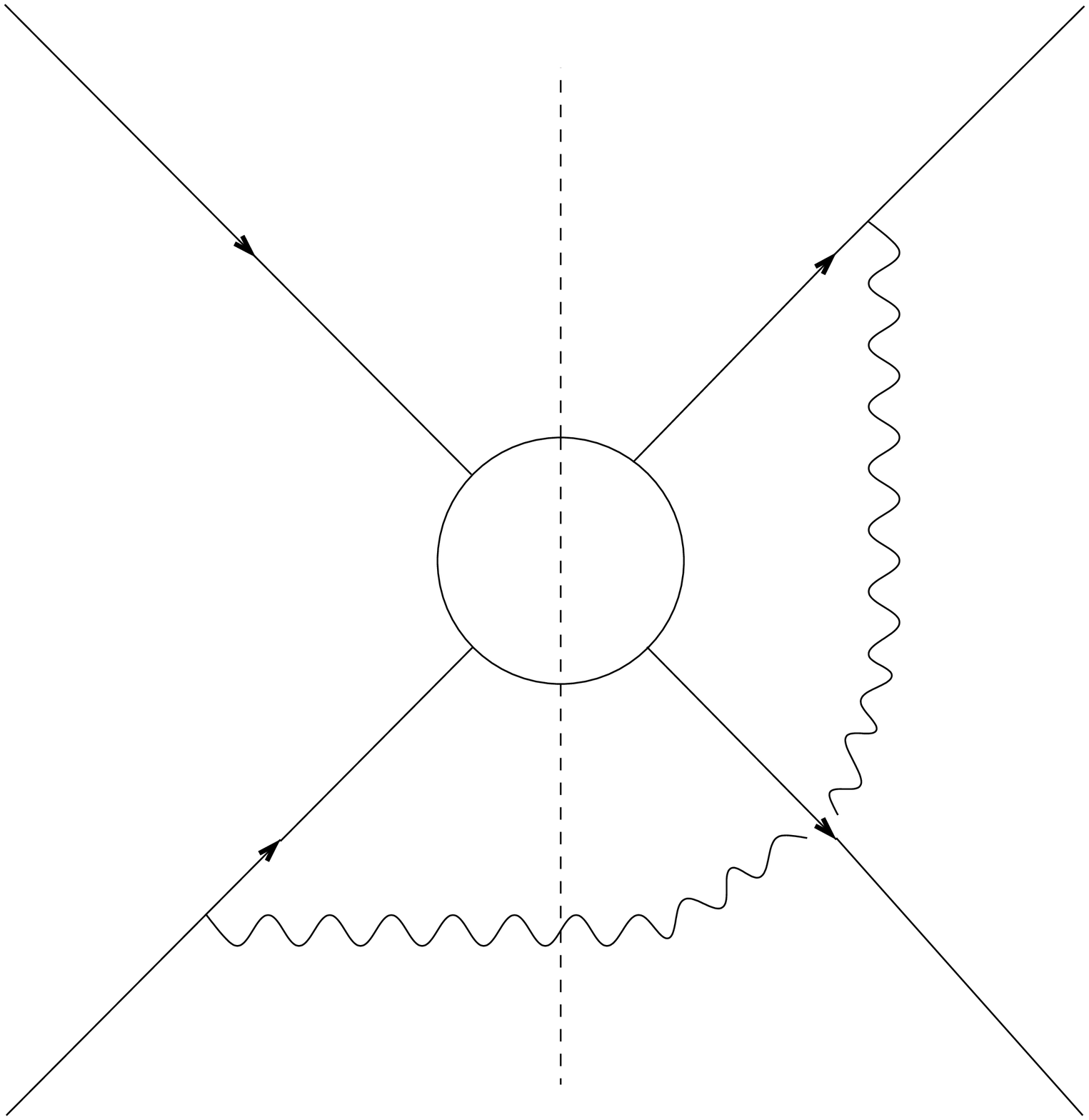,height=6cm}}
\put(11.8,3.9){$S_H^\cro\,\, S_H$}
\end{picture}
\caption{Unitarity diagrams for (a) virtual and (b) real emission
 contributions to lowest order initial state interactions 
in the Feynman gauge. Sum over gauge bosons a= $\gamma,Z,W$ and over
permutations is understood.}
\end{figure}

\end{document}